\documentclass{elsart}

\usepackage{epsfig}

\usepackage{amssymb}

\begin{document}

\begin{frontmatter}



\title{{Energy dependence of hyperon production 
in nucleus-nucleus collisions at SPS}} 

\thanks[talk]{
Corresponding author, e-mail: Domenico.Elia@ba.infn.it \\
$^1$ \hspace{-0.5mm} Permanent address: University of Udine, Udine, Italy}


{\normalsize The NA57 Collaboration}

%\author{
%}
F.~Antinori$^{l}$,
P.A.~Bacon$^{e}$,
A.~Badal{\`a}$^{g}$,
R.~Barbera$^{g}$,
A.~Belogianni$^{a}$,
A.~Bhasin$^{e}$,
I.J.~Bloodworth$^{e}$,
M.~Bombara$^{i}$,
G.E.~Bruno$^{b}$,
S.A.~Bull$^{e}$,
R.~Caliandro$^{b}$,
M.~Campbell$^{h}$,
W.~Carena$^{h}$,
N.~Carrer$^{h}$,
R.F.~Clarke$^{e}$,
A.~Dainese$^{l}$,
A.P.~de~Haas$^{s}$,
P.C.~de~Rijke$^{s}$,
D.~Di~Bari$^{b}$,
S.~Di~Liberto$^{o}$,
R.~Divi{\`a}$^{h}$,
D.~Elia$^{b,\star}$,
D.~Evans$^{e}$,
G.A.~Feofilov$^{q}$,
R.A.~Fini$^{b}$,
P.~Ganoti$^{a}$,
B.~Ghidini$^{b}$,
G.~Grella$^{p}$,
H.~Helstrup$^{d}$,
K.F.~Hetland$^{d}$,
A.K.~Holme$^{k}$,
A.~Jacholkowski$^{b}$,
G.T.~Jones$^{e}$,
P.~Jovanovic$^{e}$,
A.~Jusko$^{i}$,
R.~Kamermans$^{s}$,
J.B.~Kinson$^{e}$,
K.~Knudson$^{h}$,
A.A.~Kolozhvari$^{q}$,
V.~Kondratiev$^{q}$,
I.~Kr\'alik$^{i}$,
A.~Krav\v{c}\'akov\'a$^{j}$,
P.~Kuijer$^{s}$,
V.~Lenti$^{b}$,
R.~Lietava$^{e}$,
G.~L\o vh\o iden$^{k}$,
V.~Manzari$^{b}$,
G.~Martinsk\'a$^{j}$,
M.A.~Mazzoni$^{o}$,
F.~Meddi$^{o}$,
A.~Michalon$^{r}$,
M.~Morando$^{l}$,
E.~Nappi$^{b}$,
F.~Navach$^{b}$,
P.I.~Norman$^{e}$,
A.~Palmeri$^{g}$,
G.S.~Pappalardo$^{g}$,
B.~Pastir\v c\'ak$^{i}$,
J.~Pi\v{s}\'ut$^{f}$,
N.~Pi\v{s}\'utov\'a$^{f}$,
F.~Posa$^{b}$,
E.~Quercigh$^{l}$,
F.~Riggi$^{g}$,
D.~R\"ohrich$^{c}$,
G.~Romano$^{p}$,
K.~\v{S}afa\v{r}\'{\i}k$^{h}$,
L.~\v S\'andor$^{i}$,
E.~Schillings$^{s}$,
G.~Segato$^{l}$,
M.~Sen\'e$^{m}$,
R.~Sen\'e$^{m}$,
W.~Snoeys$^{h}$,
F.~Soramel$^{l,1}$,
M.~Spyropoulou-Stassinaki$^{a}$,
P.~Staroba$^{n}$,
T.A.~Toulina$^{q}$,
R.~Turrisi$^{l}$,
T.S.~Tveter$^{k}$,
J.~Urb\'{a}n$^{j}$,
F.~Valiev$^{q}$,
A.~van~den~Brink$^{s}$,
P.~van~de~Ven$^{s}$,
P. Vande Vyvre$^{h}$,
N.~van~Eijndhoven$^{s}$,
J.~van~Hunen$^{h}$,
A.~Vascotto$^{h}$,
T.~Vik$^{k}$,
O.~Villalobos Baillie$^{e}$,
L.~Vinogradov$^{q}$,
T.~Virgili$^{p}$,
M.F.~Votruba$^{e}$,
J.~Vrl\'{a}kov\'{a}$^{j}$ and
P.~Z\'{a}vada$^{n}$

\address[a]{\scriptsize Physics Department, University of Athens, Athens, Greece}
\vspace{-1mm}
\address[b]{\scriptsize Dipartimento IA di Fisica dell'Universit{\`a}
            e del Politecnico and INFN, Bari, Italy} 
\address[c]{\scriptsize Fysisk Institutt, Universitetet i Bergen, Bergen, Norway} 
\vspace{-1mm}
\address[d]{\scriptsize H{\o}gskolen i Bergen, Bergen, Norway} 
\address[e]{\scriptsize School of Physics and Astronomy, 
University of Birmingham, Birmingham, UK} 
\vspace{-1mm}
\address[f]{\scriptsize Comenius University, Bratislava, Slovakia} 
\address[g]{\scriptsize University of Catania and INFN, Catania, Italy} 
\vspace{-1mm}
\address[h]{\scriptsize CERN, European Laboratory for Particle Physics, Geneva,
            Switzerland} 
\vspace{-1mm}
\address[i]{\scriptsize Institute of Experimental Physics, Slovak Academy of Science,
            Ko\v{s}ice, Slovakia} 
\vspace{-1mm}
\address[j]{\scriptsize P.J. \v{S}af\'{a}rik University, Ko\v{s}ice, Slovakia} 
\vspace{-1mm}
\address[k]{\scriptsize Fysisk Institutt, Universitetet i Oslo, Oslo, Norway} 
\vspace{-1mm}
\address[l]{\scriptsize University of Padua and INFN, Padua, Italy} 
\address[m]{\scriptsize Coll\`ege de France, Paris, France} 
\address[n]{\scriptsize Institute of Physics, Academy of Sciences of the Czech Republic, Prague,
            Czech Republic} 
\address[o]{\scriptsize University ``La Sapienza'' and INFN, Rome, Italy} 
\address[p]{\scriptsize Dip. di Scienze Fisiche ``E.R. Caianiello''
            dell'Universit{\`a} and INFN, Salerno, Italy} 
\address[q]{\scriptsize State University of St. Petersburg, St. Petersburg, Russia} 
\address[r]{\scriptsize Institut de Recherches Subatomique, IN2P3/ULP, Strasbourg, France} 
\address[s]{\scriptsize Utrecht University and NIKHEF, Utrecht, The Netherlands} 
%\vspace{5mm}

\begin{abstract}
A measurement of strange baryon and antibaryon production
in Pb-Pb collisions has been carried out by the NA57
experiment at the CERN SPS, with 40
and 158 $A$ GeV/$c$ beam momentum.
Results on $\Lambda$, $\Xi$ and $\Omega$
hyperon yields at mid-rapidity in
the most central 53\% of Pb-Pb collisions at
40 $A$ GeV/$c$ are presented
and compared with those obtained
at higher energy, in the
same collision centrality range.
\\
The $\Lambda$ and $\Xi^-$ yields per unit
rapidity stay roughly constant 
while those of $\Omega^-$,
$\overline\Lambda$, $\overline\Xi^+$ and $\overline\Omega^+$
increase when 
going to the higher SPS energy.
Hyperon yields at the SPS
are compared with those
from the STAR experiment
in $\sqrt{s_{NN}}$ = 130 GeV Au-Au collisions at RHIC.
\\
\vspace{3mm}
\noindent $PACS:$ 12.38.Mh, 14.20.Jn, 25.75.Nq, 25.75.Dw
\end{abstract}

%\begin{keyword}

%\PACS 12.38.Mh \sep 25.75.Dw \sep 25.75.Nq
%\end{keyword}
\end{frontmatter}

% main text
\section{Introduction}

The CERN NA57 experiment~\cite{NA57prop} has been designed to
study the onset of the multi-strange baryon and antibaryon
enhancements in Pb-Pb collisions 
with respect to proton-induced
reactions,
first observed by the WA97 experiment
at 158 $A$ GeV/$c$ beam momentum~\cite{And99}.
NA57 has extended the WA97 measurements
over a wider centrality range and
to lower beam momentum:
data on Pb-Pb collisions have been taken at both 158 $A$ GeV/$c$
(as for WA97) and 40 $A$ GeV/$c$. 
Measurements of strangeness production
at SPS have been carried out also by the NA49
Collaboration~\cite{NA49ref}.
%programme~\cite{Afasa2002K,Afasa2002Xi,Meurer2004,Mitrovski2004}.
\\
The NA57 results on hyperon and antihyperon
production at 158 $A$ GeV/$c$~\cite{Man2002}
confirm the pattern observed
by WA97: the enhancements increase with the strangeness
content of the particle up to a factor of about 20 for
$\Omega$ hyperons.
In this paper the absolute yields 
and the inverse slopes of the 
transverse mass distributions for
$\Lambda$, $\Xi^-$ and $\Omega^-$ (and
their antiparticles) in Pb-Pb collisions at
both 40 and 158 $A$ GeV/$c$ are presented and discussed.
The results had been preliminarily
shown at the Quark Matter 2004 conference~\cite{QM2004}.
A comparison with the results
from the higher energy Au-Au collisions at RHIC is also included.

%\label{}

% \appendix

% \section{}
% \label{}

\section{The NA57 experimental layout and data sets}

\begin{figure}[htb]
\vspace{1cm}
\centering
\includegraphics[scale=0.45]{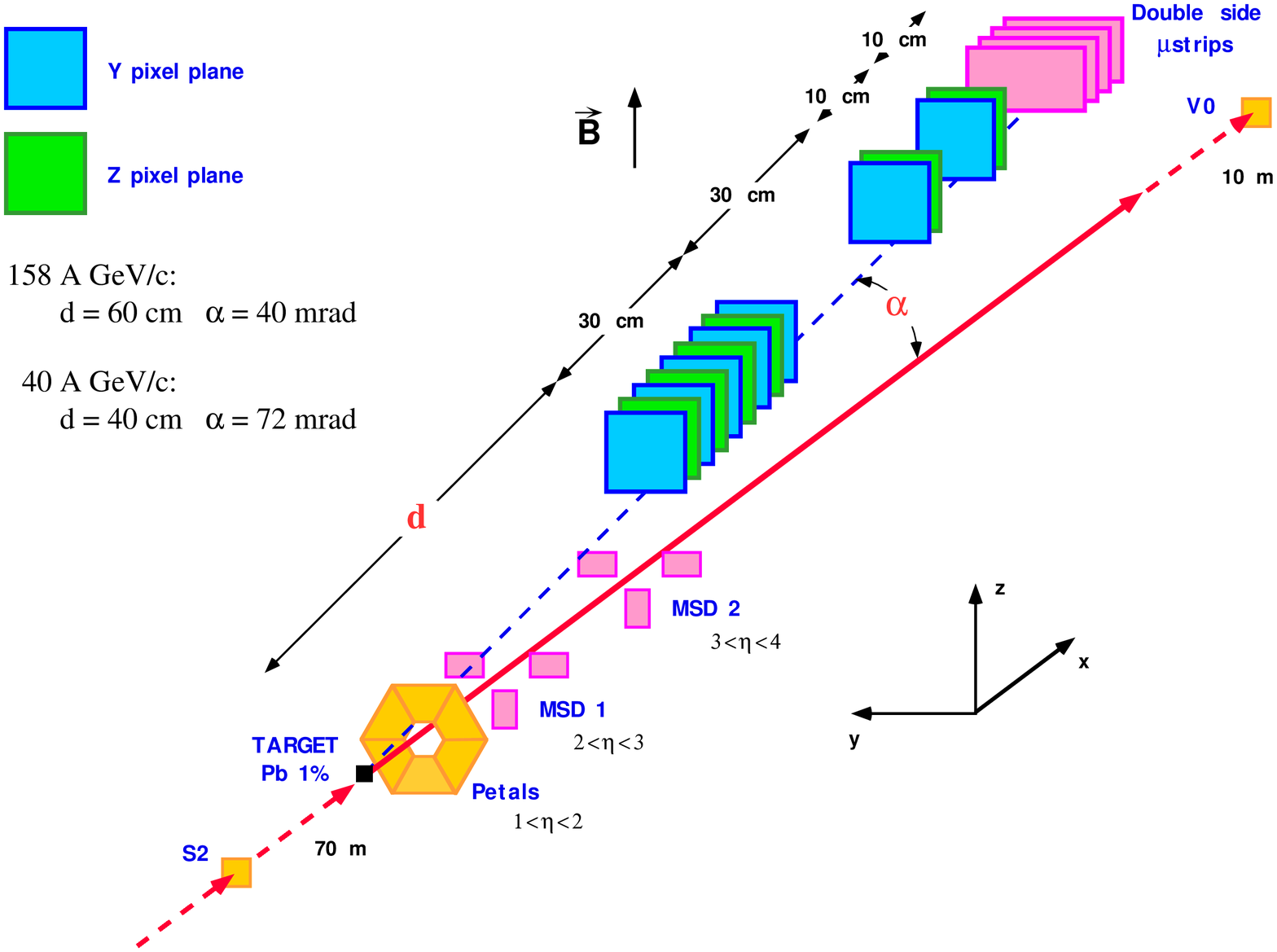}
\caption{}
\label{figlayout}
\end{figure}

The NA57 apparatus~\cite{Man99},
schematically shown in Fig.\ref{figlayout},
has been designed to detect strange and multi-strange
hyperons by reconstructing their weak decays
into final states containing charged particles only.
Tracks are measured in the silicon telescope, a
30 cm length 
array of pixel detector planes
with 5 $\times$ 5 cm$^2$ cross-section.
Additional pixel planes and 
double-sided silicon microstrip detectors, placed
behind the telescope, are used as a lever arm
to improve the resolution for high momentum tracks.
The distance from the target and
the angle of inclination of the telescope
were varied with beam momentum
in order to cover the central rapidity region.
The telescope acceptance covers about half a unit of
rapidity at central rapidity and medium transverse 
momentum\footnote[1]{The lower $p_{\rm T}$ limits
are about 0.3, 0.5 and 0.8 GeV/$c$ for $\Lambda$, $\Xi$ 
and $\Omega$ respectively.}.
\\
The centrality trigger, based on a scintillator
system (Petals) placed 10 cm downstream of the target,
selects approximately the 60\% most central
fraction of the inelastic
cross-section for Pb-Pb collisions.
The centrality of the collision
is measured using the charged particle multiplicity
sampled at central rapidity by two stations of
microstrip silicon detectors
(MSD). 
\\
The full apparatus is placed inside the
1.4 Tesla field of the GOLIATH magnet.
About 240 M and 460 M events of central Pb-Pb collisions
have been collected
at 40 and 158 $A$ GeV/$c$ respectively.

\section{Data analysis}

The following decay channels (and the
corresponding ones for antiparticles)
have been detected:
$\Lambda$ $\rightarrow$ $\pi^-{p}$,
$\Xi^-$ $\rightarrow$ $\Lambda\pi^-$ (with $\Lambda$ $\rightarrow$ $\pi^-{p}$)
and \linebreak
$\Omega^-$ $\rightarrow$ $\Lambda{K}^-$
(with $\Lambda$ $\rightarrow$ $\pi^-{p}$).
The particle selection procedure,
based on geometrical and kinematical cuts,
allows the extraction of
signals with negligible background~\cite{Man2002}.
\\
The study of the collision centrality is based on the charged
particle multiplicity sampled by the MSD in the pseudorapidity
ranges 1.9 $<$ $\eta$ $<$ 3.6 (for the 40 $A$ GeV/$c$ data) and
2 $<$ $\eta$ $<$ 4 (for the 158 $A$ GeV/$c$ 
data)\footnote[2]{The central rapidity values are 2.2 and 2.9 at 40 and
158 $A$ GeV/$c$ respectively.}, following
the analysis procedure
described in \cite{Ant2000}.
\\
For each selected particle a weight is calculated by means of
a Monte Carlo procedure based on GEANT~\cite{GEANTref} that takes into
account geometrical acceptance and reconstruction efficiency losses.
The full samples of the reconstructed $\Xi$ and $\Omega$ 
hyperons have been 
individually weighted. 
Since the method is CPU intensive, for 
the much more abundant
$\Lambda$ and $\overline\Lambda$ samples 
we only weighted a small fraction of the total
sample in order to reach a statistical error
better than the systematic error.
\\
The double differential ($y$, $m_{\rm T}$=$\sqrt{m^2+p_{\rm T}^2}$) 
invariant cross section
for each particle type
has been fitted to an exponential using the maximum likelihood
method and the parametrization \cite{Ant2004}:

\begin{equation}
\frac{1}{m_{\rm T}} \frac{d^2N}{dm_{\rm T} dy}=f(y) \hspace{1mm} \exp\left(-\frac{m_{\rm T}}{T}\right)
\hspace{65mm}.
\label{eqmtfit}
\end{equation}

\vspace{-4mm}
\noindent
We assume $f(y) = const$ (flat rapidity distribution)
within our acceptance region around central rapidity and leave
the inverse slope $T$ as a free parameter. 
\\
Yields have been
calculated as the number of particles per event extrapolated
to a common phase space
window, covering the full $p_{\rm T}$ range and one unit of rapidity
around mid-rapidity, by using the fitted slopes:

\begin{equation}
Y = \int_{m}^{\infty} dm_{\rm T} \int_{y_{cm}-0.5}^{y_{cm}+0.5} dy \frac{d^2N}{dm_{\rm T} dy}
\hspace{71mm}.
\label{eqyield}
\end{equation}

\section{Results and discussion}

Inverse slopes and yields at 40 and 158 $A$ GeV/$c$ have been
compared within a centrality range corresponding to
the most central 53\% of Pb-Pb collisions: the
average number of participants for this sample
is about 165.
\\
The $\Omega^-$ and $\overline{\Omega}^+$ samples
have been merged in the same spectrum for the lower
energy data, due to the low statistics.
The inverse slope parameters $T$ obtained with the fit
procedure and the corresponding statistical and
systematic errors are
given in Table 1.
The values of the $T$ parameter at 158 $A$ GeV/$c$
are larger than the corresponding ones
at 40 $A$ GeV/$c$
for $\Lambda$, $\overline\Lambda$ and
$\Xi^-$. Larger errors, due to the limited statistics
at lower energy,
preclude such comparisons for $\overline\Xi^+$ and $\Omega$ hyperons.
The details of the analysis 
of the transverse mass spectra at 158 $A$ GeV/$c$
are discussed in \cite{Ant2004}. A similar description
of the transverse mass spectra analysis at 40 $A$ GeV/$c$
is in preparation. 
For all the fits discussed in the
present paper we obtained a value of $\chi^2$/$dof$ below 2.
\\
These inverse slopes have been used to calculate the particle yields
given in Table 1 (according to the formula (\ref{eqyield})).
Statistical
and systematic errors are given.
The systematic errors were estimated 
by varying the analysis procedures, the 
selection criteria and the extrapolation function.
The slope extracted from the combined spectra of
$\Omega^-$ and $\overline{\Omega}^+$ has been used
for the extrapolations.
\\
Going from 40 to 158 $A$ GeV/$c$, the $\Lambda$ and $\Xi^-$ yields
at mid-rapidity are \linebreak similar while the corresponding 
antiparticle yields increase by a factor 5.
The $\Omega$ hyperon production shows a larger
increase, with about a factor 3 rise for
the particle and more than a factor 7 for the antiparticle.
These results can be understood as due to
a higher
baryon density in the collision fireball 
at lower energy.

\vspace{3mm}
{Table 1. Inverse slope parameters and yields for the
most central 53\% of Pb-Pb collisions
at 40 and 158 $A$ GeV/$c$.
Statistical (first) and systematic (second) errors are also quoted.}
%\begin{table}
%\footnotesize
\scriptsize
\begin{center}
\begin{tabular}{|c|c|c|c|c|} \hline
{} & \multicolumn{2}{|c|}{{\bf\em T} (MeV)} 
& \multicolumn{2}{|c|}{{\bf Yield}} \\
{} & \multicolumn{1}{|c}{40 $A$ GeV/$c$} & \multicolumn{1}{c|}{158 $A$ GeV/$c$}
   & \multicolumn{1}{|c}{40 $A$ GeV/$c$} & \multicolumn{1}{c|}{158 $A$ GeV/$c$} 
   \\ \hline\hline
{$\Lambda$}          & {261 $\pm$ 4 $\pm$ 26} & {289 $\pm$ 7 $\pm$ 29}
                     & {7.18 $\pm$ 0.13 $\pm$ 0.72} & {7.84 $\pm$ 0.21 $\pm$ 0.78} 
		     \\ \cline{2-5}
{$\overline\Lambda$} & {263 $\pm$ 6 $\pm$ 26} & {287 $\pm$ 6 $\pm$ 29}
                     & {0.188 $\pm$ 0.005 $\pm$ 0.019} & {1.17 $\pm$ 0.03 $\pm$ 0.12} 
		     \\ \cline{2-5}
{$\Xi^-$}            & {228 $\pm$ 12 $\pm$ 23} & {297 $\pm$ 5 $\pm$ 30}
                     & {0.635 $\pm$ 0.041 $\pm$ 0.064} & {0.852 $\pm$ 0.015 $\pm$ 0.085} 
		     \\ \cline{2-5}
{$\overline\Xi^+$}   & {308 $\pm$ 63 $\pm$ 31} & {316 $\pm$ 11 $\pm$ 30}
                     & {0.044 $\pm$ 0.008 $\pm$ 0.004} & {0.209 $\pm$ 0.007 $\pm$ 0.021} 
		     \\ \cline{2-5}
{$\Omega^-$ + $\overline{\Omega}^+$}  
                     & {368 $\pm$ 120 $\pm$ 40} & {271 $\pm$ 16 $\pm$ 27}
                     & {} & {} 
		     \\ \cline{2-5}
{$\Omega^-$}         & {} & {264 $\pm$ 19 $\pm$ 27}
                     & {0.039 $\pm$ 0.014 $\pm$ 0.004} & {0.118 $\pm$ 0.011 $\pm$ 0.012} 
		     \\ \cline{2-5}
{$\overline{\Omega}^+$} 
                     & {} & {284 $\pm$ 28 $\pm$ 27}
             	     & {0.007 $\pm$ 0.003 $\pm$ 0.001} & {0.054 $\pm$ 0.007 $\pm$ 0.005} 
		     \\ \hline
\end{tabular}
\end{center}
\normalsize
%{Table 1. Inverse slope parameters and yields in Pb-Pb collisions
%at 40 and 158 $A$ GeV/$c$.
%Statistical (first) and systematic (second) errors also are quoted.}
\vspace{3mm}

\begin{figure}[htb]
\includegraphics[scale=0.55]{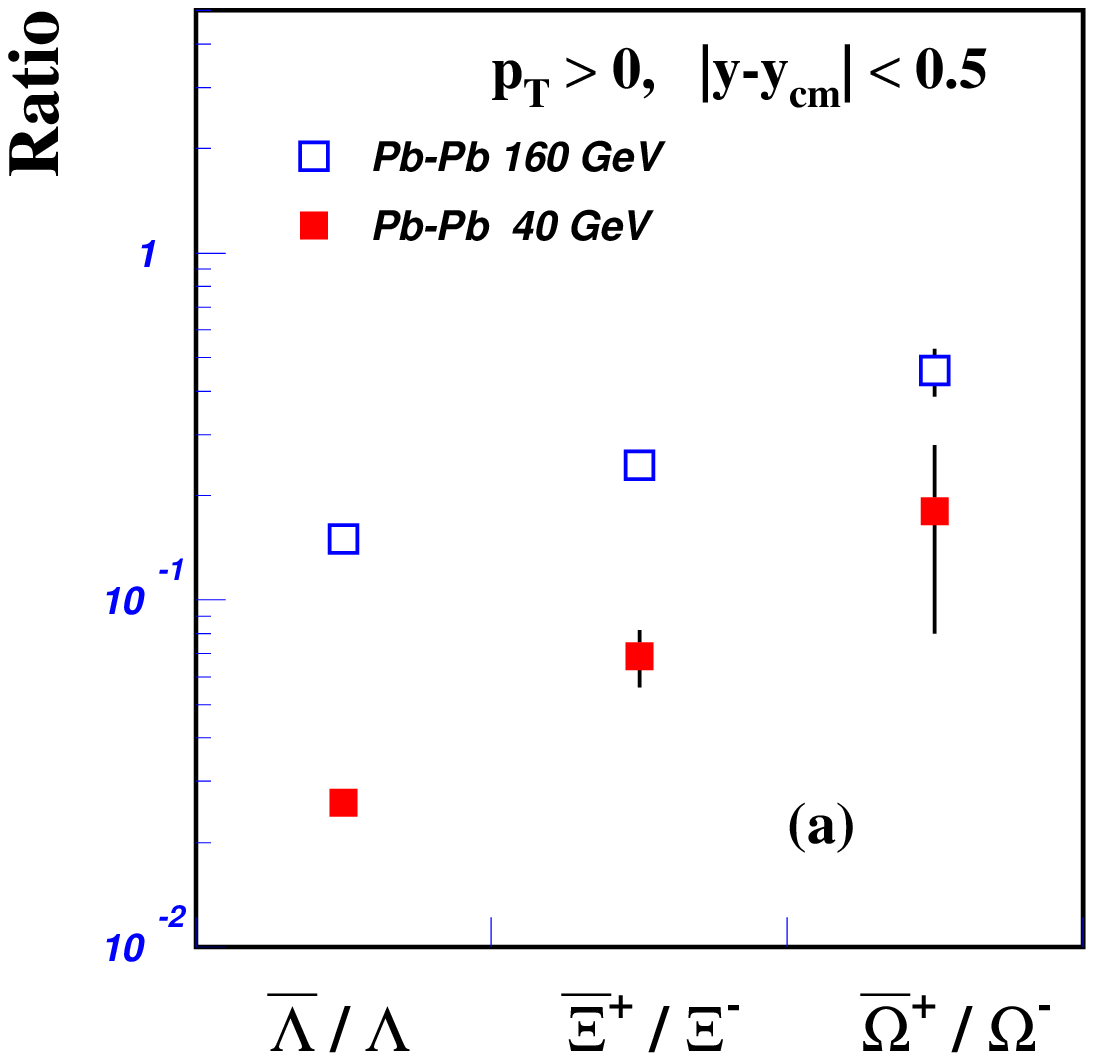}
\hspace{\fill}
\begin{minipage}[t]{110mm}
\includegraphics[scale=0.55]{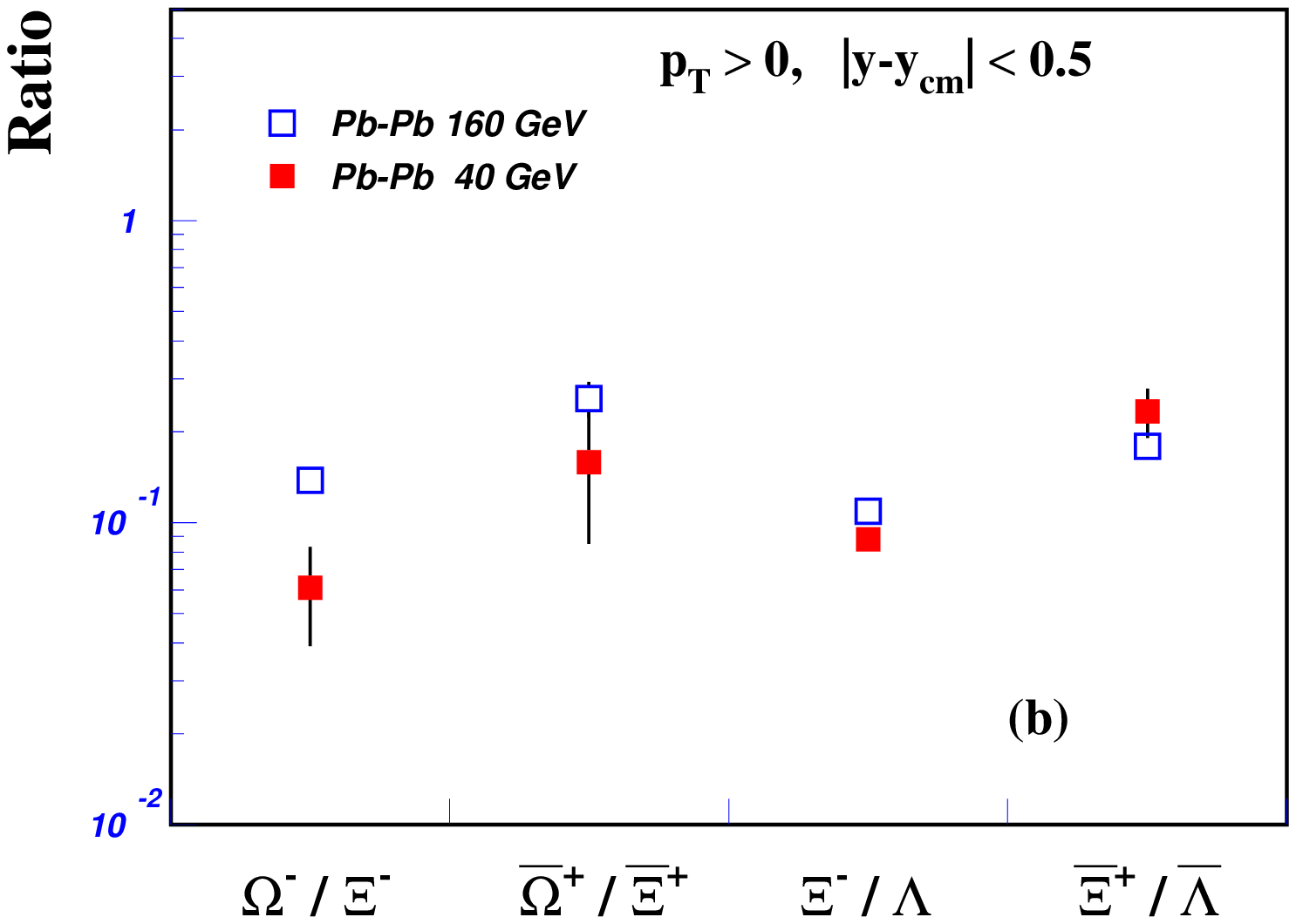}
\end{minipage}
\caption{}
\label{figrat}
\end{figure}

From the above yields, production ratios for 
different particle species 
are calculated and shown in Table 2.
Fig.\ref{figrat}
shows these particle ratios, keeping
the antiparticle to particle (a) 
separate from the mixed ones (b).
Only the statistical errors are shown. We
estimate the systematic errors to be of the 
order of 5\% for the antiparticle to particle ratios
and of the order of 10\% for the mixed ratios.

\vspace{3mm}
{Table 2. Particle ratios in the most central 53\% of
Pb-Pb collisions at 40 and 158 $A$ GeV/$c$.
Quoted errors are statistical only.}
\scriptsize
\begin{center}
\begin{tabular}{|c|c|c|} \hline
{} & \multicolumn{2}{|c|}{\bf Particle ratio} \\
{} & \multicolumn{1}{|c}{40 $A$ GeV/$c$} & \multicolumn{1}{c|}{158 $A$ GeV/$c$} \\ \hline\hline
{$\overline\Lambda$ / $\Lambda$} 
	& {0.026 $\pm$ 0.001} & {0.149 $\pm$ 0.006} \\ \cline{2-3}
{$\overline\Xi^+$ / $\Xi^-$}
	& {0.069 $\pm$ 0.013} & {0.245 $\pm$ 0.009} \\ \cline{2-3}
{$\overline{\Omega}^+$ / $\Omega^-$} 
	& {0.18 $\pm$ 0.10} & {0.458 $\pm$ 0.073} \\ \hline
{$\Omega^-$ / $\Xi^-$} 
	& {0.061 $\pm$ 0.022} & {0.138 $\pm$ 0.013} \\ \cline{2-3}
{$\overline{\Omega}^+$ / $\overline\Xi^+$} 
	& {0.159 $\pm$ 0.074} & {0.258 $\pm$ 0.035} \\ \cline{2-3}
{$\Xi^-$ / $\Lambda$} 
	& {0.088 $\pm$ 0.006} & {0.109 $\pm$ 0.003} \\ \cline{2-3}
{$\overline\Xi^+$ / $\overline\Lambda$} 
	& {0.234 $\pm$ 0.043} & {0.179 $\pm$ 0.008} \\ \hline
\end{tabular}
\end{center}
\normalsize
%{Table 2. Particle ratios in Pb-Pb collisions at 40 and 158 $A$ GeV/$c$.
%Quoted errors are statistical only.}
\vspace{3mm}

\begin{figure}[htb]
\includegraphics[scale=0.55]{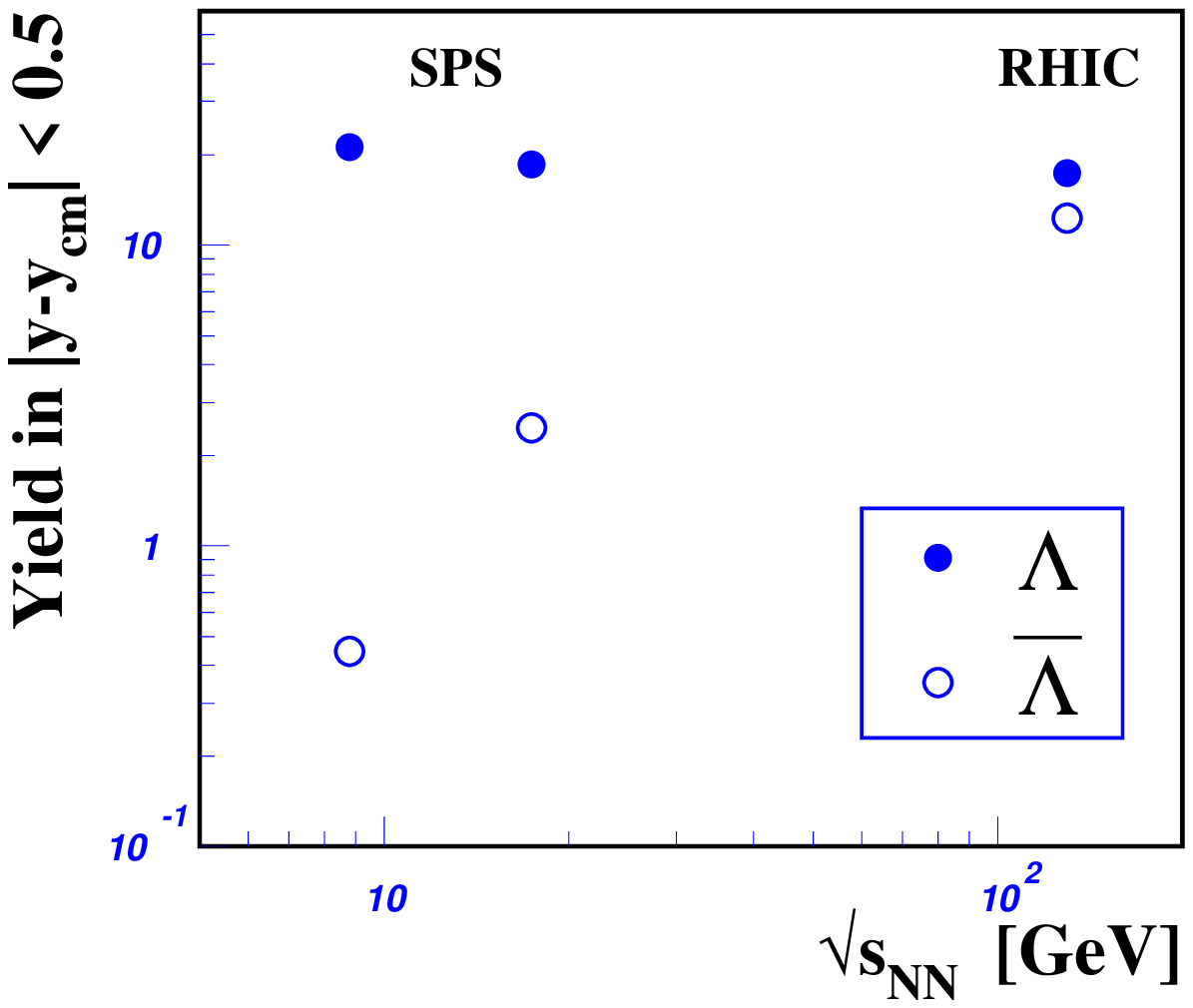}
\hspace{\fill}
\begin{minipage}[t]{110mm}
\includegraphics[scale=0.55]{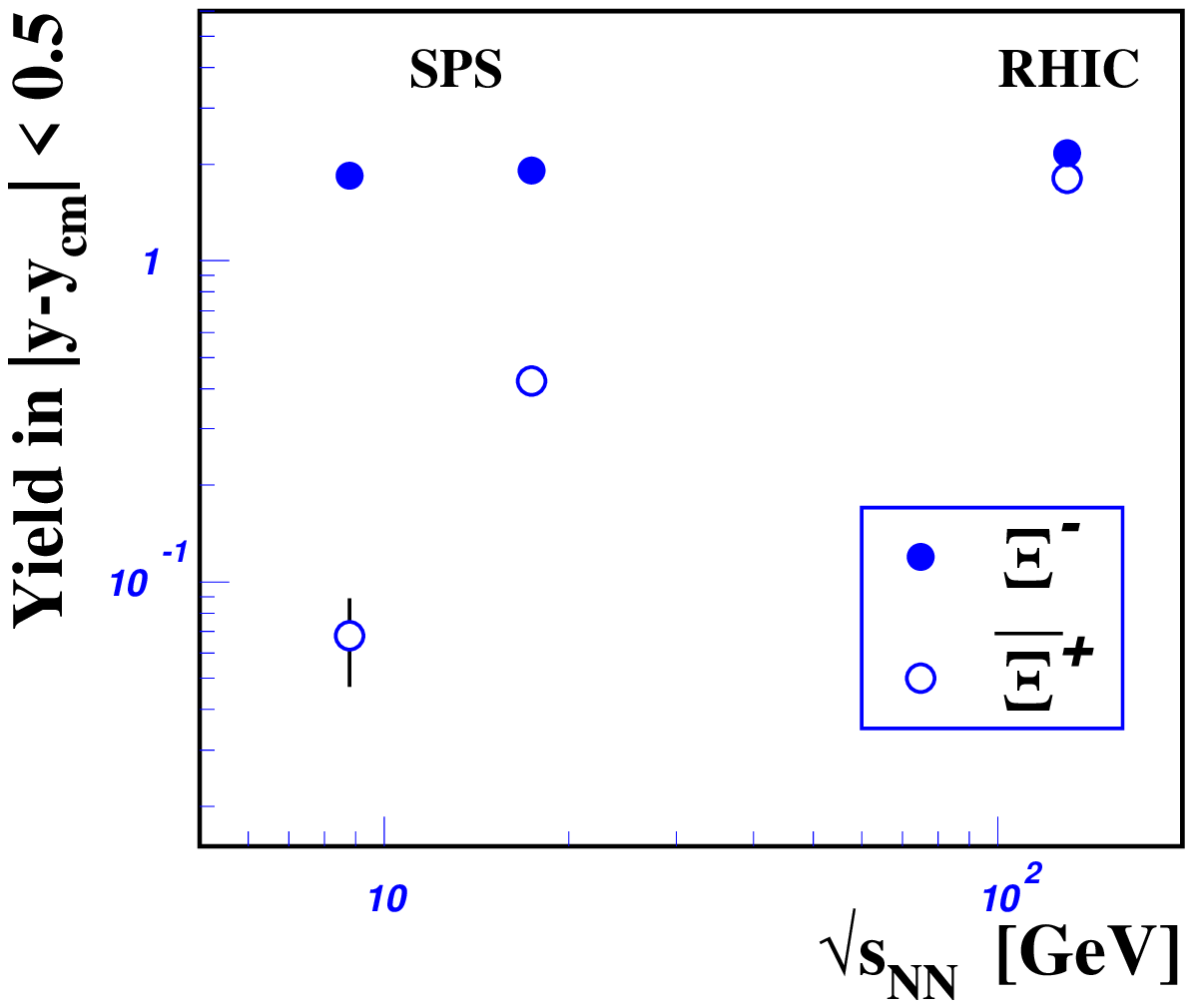}
\end{minipage}
\\
\begin{center} 
\includegraphics[scale=0.55]{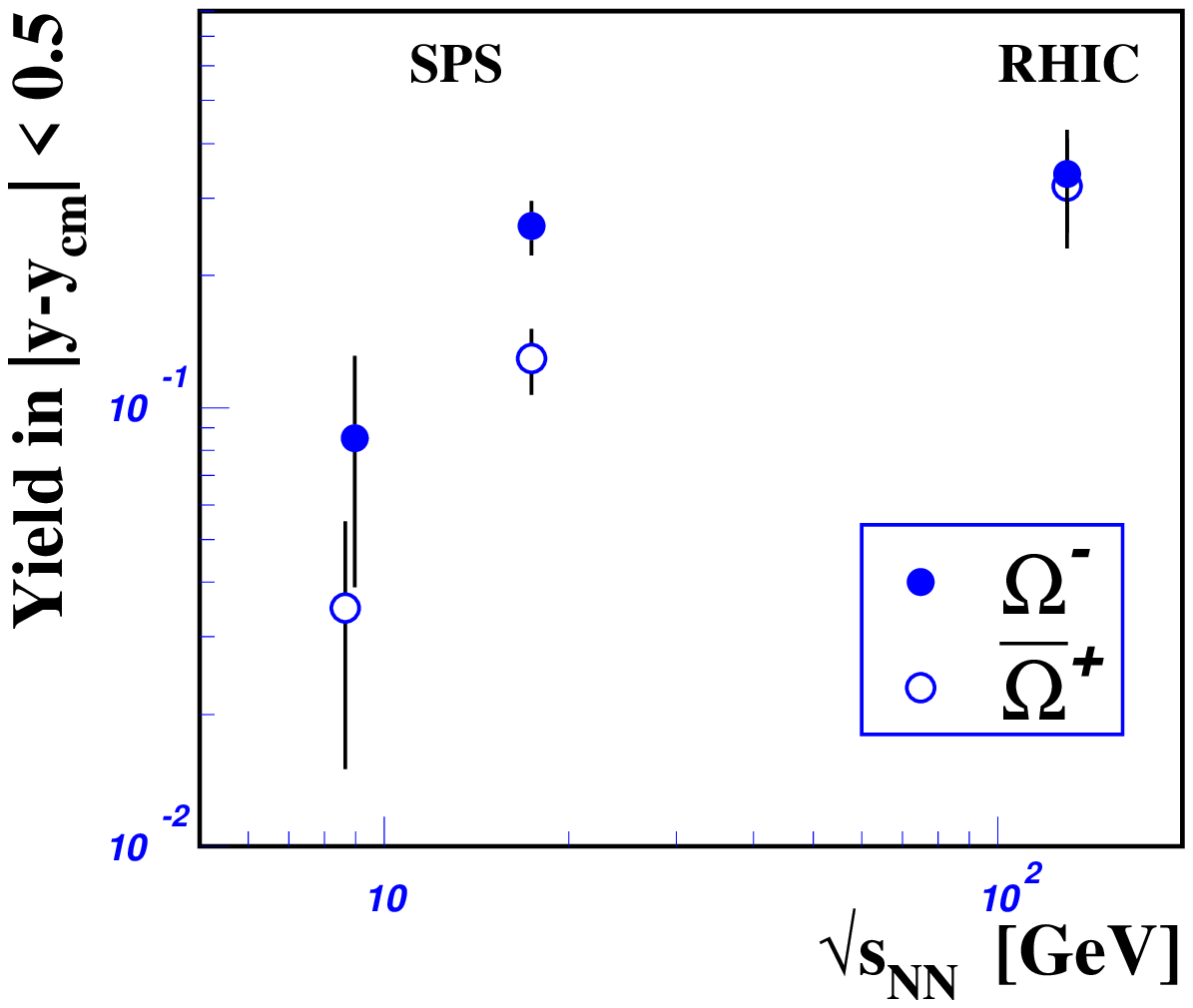}
\end{center} 
\caption{}
\label{figyiecomp}
\end{figure}

When going from 40 to 158 $A$ GeV/$c$ the antihyperon to
hyperon ratios go up, the effect being largest for $\Lambda$.
As for mixed ratios we do not see any significant variation,
except for 
$\Omega^-$/ $\Xi^-$ which increases by more than a factor 2.
\\
Results on hyperon production
at mid-rapidity
in $\sqrt{s_{NN}}$ = 130 GeV Au-Au collisions
from the STAR experiment at RHIC are also 
available~\cite{STARrefyie}.
%available~\cite{Adler2002,Adams2003,Suire2002}.
For comparison we selected our data in the same
centrality range used for STAR results
(most central 5\%, 10\%, 11\%
collisions for $\Lambda$, $\Xi$ and $\Omega$ respectively).
%In Fig.\ref{figyiecomp} the NA57 hyperon yields per
The corresponding yields per
unit rapidity at 40 $A$ GeV/$c$
($\sqrt{s_{NN}}$ = 8.8 GeV) and 158 $A$ GeV/$c$
($\sqrt{s_{NN}}$ = 17.3 GeV) are given in Table 3
and compared with those
from STAR in Fig.\ref{figyiecomp}. 

\vspace{3mm}
{Table 3. Hyperon yields in the
most central 5\%, 10\% and 11\% of Pb-Pb collisions
for $\Lambda$, $\Xi$ and $\Omega$
respectively,
at 40 and 158 $A$ GeV/$c$.
Statistical (first) and systematic (second) errors are also quoted.}
%\begin{table}
%\footnotesize
\scriptsize
\begin{center}
\begin{tabular}{|c|c|c|} \hline
{\hspace{1.3cm}}
& \multicolumn{2}{|c|}{{\bf Yield}} \\
{} & \multicolumn{1}{|c}{40 $A$ GeV/$c$} & \multicolumn{1}{c|}{158 $A$ GeV/$c$}
   \\ \hline\hline
{$\Lambda$}         
                     & {21.1 $\pm$ 0.8 $\pm$ 2.1} & {18.5 $\pm$ 1.1 $\pm$ 1.9} 
		     \\ \cline{2-3}
{$\overline\Lambda$}
                     & {0.44 $\pm$ 0.03 $\pm$ 0.04} & {2.47 $\pm$ 0.14 $\pm$ 0.25} 
		     \\ \cline{2-3}
{$\Xi^-$}            
                     & {1.84 $\pm$ 0.16 $\pm$ 0.18} & {1.91 $\pm$ 0.05 $\pm$ 0.19} 
		     \\ \cline{2-3}
{$\overline\Xi^+$}   
                     & {0.068 $\pm$ 0.021 $\pm$ 0.007} & {0.422 $\pm$ 0.023 $\pm$ 0.042} 
		     \\ \cline{2-3}
{$\Omega^-$}         
                     & {0.085 $\pm$ 0.046 $\pm$ 0.009} & {0.259 $\pm$ 0.037 $\pm$ 0.026} 
		     \\ \cline{2-3}
{$\overline{\Omega}^+$} 
                     
             	     & {0.035 $\pm$ 0.020 $\pm$ 0.004} & {0.129 $\pm$ 0.022 $\pm$ 0.013} 
		     \\ \hline
\end{tabular}
\end{center}
\normalsize
\vspace{4mm}

\begin{figure}[htb]
\centering
\includegraphics[scale=0.53]{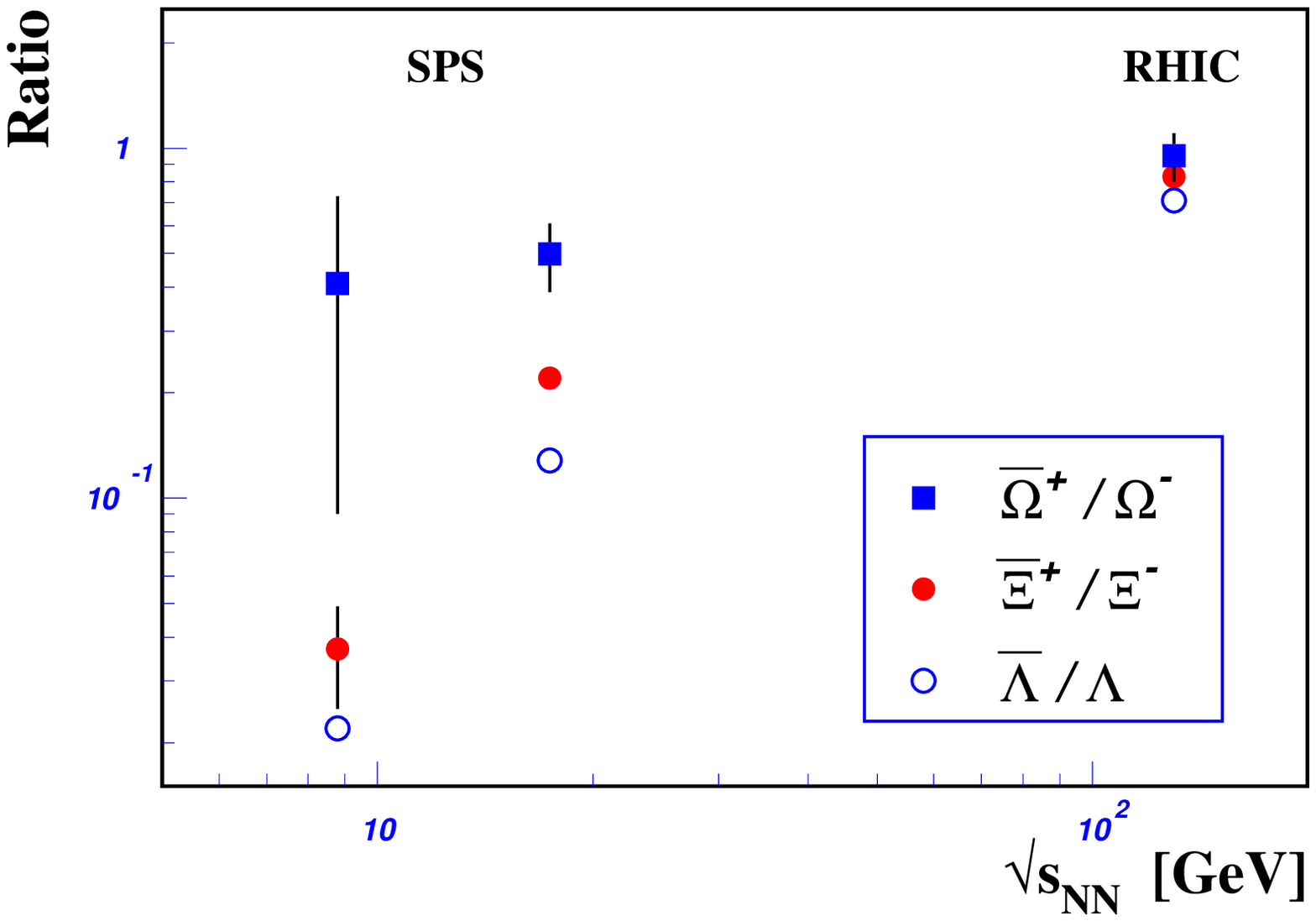}
\caption{}
\label{figratcomp}
\end{figure}

The $\Lambda$ and $\Xi^-$ yields
do not vary much from SPS to RHIC energies.
A clear energy dependence
is seen for the three antihyperon yields.
\\
The antihyperon to hyperon ratios are plotted in 
Fig.\ref{figratcomp} as a function of
$\sqrt{s_{NN}}$ from SPS to RHIC~\cite{Adams2003PL}. 
At RHIC the ratios increase
with increasing strangeness content of the hyperon,
as already seen at SPS energies.
All three ratios also increase as a function
of the energy,
the dependence being weaker for
particles with higher strangeness.

\section{Conclusions}

Results from NA57 on hyperon production in
Pb-Pb collisions at 40 and 158 $A$ GeV/$c$
beam momentum
have been presented and compared.
The $\Lambda$ and $\Xi^-$ yields per unit
rapidity stay roughly constant 
while those of $\Omega^-$,
$\overline\Lambda$, $\overline\Xi^+$ and $\overline\Omega^+$
increase when 
going to the higher SPS energy.
\\
Comparison with higher energy results from the STAR
experiment at RHIC
confirms the same trend, with hyperons
showing a weaker energy dependence than 
antihyperons. The antihyperon to hyperon ratios
increase with the \linebreak energy, 
with a stronger dependence
for particles with smaller strangeness content.
Such a pattern is consistent with
a decrease of baryon density in the central region
with increasing energy.

%\newpage
\vspace{2.5cm}

\noindent{\bf Figure caption}
\vspace{0.5cm}

Fig. 1. The NA57 experimental layout for Pb-Pb collisions.
\\

Fig. 2. Antiparticle to particle ratios (a) and
mixed particle ratios (b) in the most central 53\% of 
Pb-Pb collisions at 40 and 158 $A$ GeV/$c$.
Errors shown are statistical only.
\\

Fig. 3. Hyperon yields at central rapidity at SPS and RHIC energies.
The selected data samples correspond to the most central 5\%, 10\% and 11\%
collisions for $\Lambda$, $\Xi$ and $\Omega$ respectively.
Errors shown are statistical only.
\\

Fig. 4. Comparison of antihyperon to hyperon
ratios at SPS and RHIC energies.
The selected data samples correspond to the most central 5\%, 10\% and 11\%
collisions for $\Lambda$, $\Xi$ and $\Omega$ respectively.
Errors shown are statistical only.
\\

\newpage

\vspace{3cm}

\end{document}